\documentclass{elsart}
\usepackage{epsfig}
\usepackage[tablesright]{rotating}
\usepackage{longtable}
\usepackage{lscape}
\usepackage{rotating}
\begin{document}
\runauthor{M. Biswas et. al}
\begin{frontmatter}
\bibliographystyle{unsrt}
\title{ The study of threshold behaviour of effective potential for $^{6}$Li+$^{58,64}$Ni
systems}
\author[SINP]{M. Biswas},
\author[SINP]{Subinit Roy \thanksref{email}},
\author[SINP]{M. Sinha},
\author[SINP]{M.K. Pradhan},
\author[SINP]{A. Mukherjee},
\author[SINP]{P. Basu},
\author[SINP]{H. Majumdar},
\author[BARC]{K. Ramachandran},
\author[BARC]{A. Shrivastava}
\thanks[email]{Corresponding author. E-mail: subinit.roy@saha.ac.in}
\address [SINP]{Saha Institute of Nuclear Physics, I/AF Bidhan Nagar, Kolkata - 700064, India.}
\address [BARC]{Nuclear Physics Division, Bhabha Atomic Research Centre, Mumbai-400085, India.}

\begin{abstract}
The elastic scattering for $^6$Li+$^{64}$Ni system was measured in the bombarding energy range of 
13 MeV $\leq$ $E_{lab}$ $\leq$ 26 MeV. A phenomenological optical model analysis was performed and
the behaviour of the surface strengths of the potential components with decreasing energy was extracted.
A further analysis of the measured angular distributions, along with
the existing data for $^6$Li+$^{58}$Ni, was performed with two different model potentials - one with 
the folded potential normalized with a complex factor (OMP1) and the other with a {\it hybrid} potential 
composed of a renormalized folded real and a phenomenological imaginary (OMP2) potential components - were 
used in the calculation. All the model potentials predict similar energy dependent behaviour for the interaction 
potential around the barrier. The observed energy dependence of the strengths of the real and imaginary potentials 
corroborate with the dispersion relation prediction for both the $^6$Li+$^{64}$Ni and $^6$Li+$^{58}$Ni systems. 
Though the evidence of breakup is distinct in the energy variation of the potential strengths, close to the barrier 
the variation is more in the line of conventional threshold anomaly. Also the threshold behaviour of the interaction 
potential does not indicate any distinct isotopic dependence. 
\end{abstract}


\end{frontmatter}

\section{Introduction}
	 
Loosely bound nuclei are liable to breakup in a nuclear collision process. 
The breaking up of the nucleus can, subsequently, affect the other 
nuclear processes in a manner not observed in collisions involving only tightly bound 
nuclei. Extensive investigations, both experimentally and theoretically, are
being pursued in recent years to understand the effect of breakup on elastic
scattering \cite{kel1,signo,so1,figu1,cama,mor1,pak1}. One of the primary motivations
is to probe the influence of break up process on the threshold behaviour of optical 
potential at near the Coulomb barrier energies.

The conventional threshold behaviour of the optical potential for tightly bound nuclei shows
a rapid rise in the strength of the real potential in the vicinity of the Coulomb barrier
with a sharp fall in the strength of the imaginary potential in the same energy range. It has been 
demonstrated that the coupling of relative motion to other reaction channels like 
inelastic excitations, transfers to the bound states contributes an attractive real polarization
potential which in turn enhances the real strength at energies around the barrier \cite{satch1,satch2}.
The rapid decrease of fusion and other of the reaction processes with decreasing energy
induces the sharp fall in the strength of the imaginary potential.
The observed threshold behaviours of the potential components for the tightly bound nuclei are demonstrated
to be connected by a dispersion relation, a manifestation of causality in nuclear collision process
\cite{satch2}.

On the other hand, the dynamical effect of coupling to break up channels of weakly bound projcetiles is to 
introduce a repulsive real polarization potential \cite{mahaux,saku1}. 
At above barrier energies the effect of breakup of the projectile was demonstrated through 
the reduction needed in the strength of the folded real potential describing 
the elastic scattering of projectiles like $^{6,7}$Li and $^{9}$Be \cite{satch1}. With the bombarding energy
decreasing for the scattering of $^6$Li projectile (breakup threshold = 1.67 MeV) from heavy targets,  
the real potential strength exhibits a decline around the Coulomb barrier energy while the imaginary 
potential exhibits a rise in the surface strength \cite{kel1,hussein}. 
Attempt has been made to understand this behaviour in terms of dispersion relation 
\cite{hussein} and the observed threshold behaviour is termed as breakup threshold anomaly (BTA). 
For targets like $^{208}$Pb, due to strong Coulomb field, the breakup reaction channels do not close down 
rapidly near the barrier or at sub-barrier energies. Experimental observations suggest that the breakup
cross section is even larger than the fusion cross section in the vicinity of the barrier \cite{hinde,aguilera,wu,woollis}. As the coupling to breakup process contributes a repulsive dynamical polarization potential, the 
normal threshold anomaly (TA) in the real potential arising out of the attractive real polarization contribution
of the conventional couplings, may disappear. The effective real potential strength may even show a declining trend with 
with further decrease in energy if the breakup coupling dominates over the conventional couplings at near barrier energies.
Recently, So, {\it et al.} \cite{so2} have shown, based 
on the extended optical model approach, that the real and imaginary polarization potential components for scattering 
of $^6$Li from $^{208}$Pb, when decomposed into direct and fusion contributions, satisfy the dispersion relation 
separately. The fusion part in this case showing a conventional threshold behaviour while the direct part exhibits a 
smooth and relatively weak energy dependence.
For $^7$Li, a more strongly bound projectile (b.u. threshold = 2.43 MeV) compared to $^6$Li, the optical potential 
describing the scattering from heavy targets shows up a threshold behaviour 
like the strongly bound projectiles \cite{kel1,martel} and the energy variations of the real and imaginary components 
can be connected by the dispersion relation prediction. 

Scattering of the weakly bound nuclei on lighter targets throws up a more complicated situation. 
The analyses of elastic scattering of $^{6,7}$Li from $^{28}$Si \cite{pak1,pak2} show that both these
projectiles, despite having different breakup threshold, exhibit similar trend in energy dependence of 
optical potential components. The imaginary potential for $^6$Li+$^{28}$Si does not show the characteristic 
rise on approaching the barrier, rather it decreases after a small rise in strength just above the barrier. 
The real potential on the other hand is energy independent with a slight tendency to decrease as energy is 
decreased. Unlike the observations of scattering from lighter target nuclei, F.A. Souza, {\it et al.} \cite{souza} 
have found that the optical potential describing the elastic scattering of $^6$Li and $^7$Li from medium mass 
$^{59}$Co exhibits a behaviour similar to that of the more tightly bound systems indicating the existence of a 
conventional threshold anomaly. However, as pointed out by the authors, looking at the energy dependence of the 
real and imaginary strengths of $^6$Li+$^{59}$Co system carefully one can also categorize the threshold behaviour 
as breakup threshold anomaly (BTA). On approaching the barrier, the real strength shows a weak declining trend. 
The surface strength of the imaginary potential rises in the same energy region before falling again with decreasing 
energy. No attempt has been made to establish the dispersion relation connectivity between the observed behaviours. 
Further investigation of $^6$Li scattering from medium mass targets is, therefore, necessary to identify the general 
behaviour of the interaction potential around the barrier. In general, for medium and light mass targets with reduced 
strength of the Coulomb field, the Coulomb induced breakup cross section is expected to decrease near the barrier. Thus 
the open question is how far the breakup of loosely bound projectiles, in the presence of conventional couplings, determines 
the threshold behaviour of interaction potentials for medium or light mass targets? The answer to this question will 
also address to the problem of target mass dependence of threshold anomaly for loosely bound projectiles. 

We report, in this context, a systematic investigation of elastic scattering of $^6$Li from $^{58,64}$Ni 
targets at bombarding energies around the barrier. The isotopes of Ni have very similar deformation values,
$\beta_2$=0.17 and 0.18 respectively for $^{58}$Ni and $^{64}$Ni. It was demonstrated by Keeley {\it et al.} 
\cite{kel2} that both these isotopes exhibit similar threshold behaviour of potential for scattering with
$^{16}$O indicating that the threshold anomaly developed predominantly from inelastic couplings. We intend
to investigate the energy dependence of the optical potential for the projectile $^6$Li scattered from these two 
Ni isotopes. The primary interest is to see the manifestation of coupling to the continuum on the threshold behaviour
of the inetraction potential in the presence of strong inelastic couplings. The elastic scattering angular distributions 
for $^6$Li+$^{64}$Ni have been measured at bombarding energies around the barrier to compare with the existing data of 
$^6$Li+$^{58}$Ni \cite{pfeiffer}. The experimental details have been given in Section 2. The model calculation and the 
analysis have been presented in Section 3, followed by the results and discussion in Section 4.

\section{Experimental Details}

The experiment was carried out at 14UD TIFR/BARC Pelletron Facility in Mumbai, India. Self-supporting targets 
were prepared by electron gun evaporation technique from 99\% enriched metallic $^{64}$Ni isotope. Two different
targets of thickness 61$\mu$g/cm$^2$ and 376$\mu$g/cm$^2$ respectively, were used for the experiment. The 
targets were bombarded with $^6$Li beam from the pelletron at energies of 13, 14, 17, 19 and 26 MeV. 
The beam current during the experiment was varied from 1 to 7 pnA. The current was
measured using a Current Integrator, the output of which was fed into a CAMAC Scaler to obtain the integrated
charge. The elastically scattered particles were detected with conventional telescope arrangement. Two silicon surface 
barrier detector telescopes were used in the two parts of the experiment. In the first part the telescopes were set with 
25$\mu$ and 15$\mu$ thick $\Delta$E detectors and 300$\mu$ thick E detectors to detect the scattered particles. In 
the second part the $\Delta$E detectors of thickness 15$\mu$ and 10$\mu$ were used and these were followed by 
3mm and 500$\mu$ E detectors respectively. The telescopes were placed on a rotatable arm at an angular separartion of 
10$^o$. The solid angles subtended by the telescopes at the target centre were 
0.076, 0.17, 0.06 and 0.11 msr respectively. The angular opening corresponding to these solid angle values were in 
the range of 0.5$^o$ to 0.8$^o$. Two monitor detectors of thickness 3mm  were mounted at $\pm$15$^o$ about the beam axis.
at a distance of 41.7cm in a 1m diameter scattering chamber. These two detectors were used to monitor the beam position 
and also for the purpose of normalization. Calibration runs were taken with a standard Bi target after each energy change. 
Using the Rutherford scattering between $^6$Li and $^{209}$Bi at these energies, the detector solid angles and the relative 
normalization between the telescopes were determined from the calibration runs. The statistical error in the data is less 
than 1\% in the forward angles and a maximum of 16\% in the backwards angles. The overall error in the data varied from 
5\% to 17\%. The data were recorded using the Linux based data acquisition system LAMPS \cite{amber}. The measured elastic 
cross sections with respect to the Rutherford cross sections are shown in Fig.1 along with the model (see next section) 
calculations.

\section{Analysis}
Optical model (OM) analysis was carried out using three different model potentials. Firstly we performed the optical model
analysis of our $^6$Li+$^{64}$Ni data using the parametric Woods Saxon (WS) forms for both the real and imaginary potentials.
Subsequently, the new data of $^6$Li+$^{64}$Ni and the existing data of $^6$Li+$^{58}$Ni \cite{pfeiffer} were further 
analysed with two other model potentials, namely, (i) the renormalized folded real and imaginary potentials, and (ii) a 
'hybrid' potential consisting of a renormalized folded real and a parametric Woods Saxon imaginary components. The use 
of different model potentials was intended to make the extraction of energy dependence of the surface strengths of the 
potential components as much model independent as possible.

\subsection{Phenomenological OM analysis of $^6$Li+$^{64}$Ni data}
The phenomenological optical model potential used to describe the elastic angular distributions at each energy had the
following form

\begin{equation}
U_{OM}(R) = V(R; V_o,R_o,a_o) + i [W_F(R; W_o,R_w,a_w) + W_D(R; W_s,R_s,a_s)]
\end{equation}

where V(R) denotes the volume type WS real potential, $W_F$(R) is a volume type WS imaginary potential to simulate 
the fusion after penetration of the barrier and $W_D$(R) is a derivative type WS imginary potential to account for 
the absorption due to reactions occuring at the surface. 
 
The fitting procedure to obtain the parameters of the best fit potential can be summarized as follows. The search 
code ECIS94 \cite{raynal} was used to perform the model calculations.
The volume imaginary potential $W_F$ was kept fixed for all the energies with the parameters set at $W_o$=50.0 MeV, 
$r_w$=1.0 fm and $a_w$=0.25 fm. At the highest energy, the real potential parameters for $^6$Li+$^{58}$Ni from 
Ref.\cite{pfeiffer} were used as the starting parameters. Keeping the real radius fixed, searches were performed 
over the remaining five free parameters, {\it viz.}, $V_o$, $a_o$, $W_s$, $R_s$ and $a_s$. Subsequently, changing 
the real radius in steps, same search was repeated again to obtain the best fit parameters with minimum $\chi^2$ 
value. The final set of best fit parameters for 26 MeV corresponding to minimum $\chi^2$/N value are given in Table 1. 
For the other incident energies the same search procedure was followed with the best fit parameters of 26 MeV as the 
starting parameter set. It is to emphasized that the resultant geometry parameters along with the strengths of the 
potential components are found to be energy dependent. For instance, the radius values of both the real and imaginary 
components increases with decreasing energy. The best fit parameters, the minimum $\chi^2$/N values and the corresponding 
reaction cross sections $\sigma_{reac}$ have been given in Table 1. The fits are shown by solid lines in Fig.1.

To probe the real and imaginary potentials as a function of energy in the vicinity of the barrier, 
it is important to identify the radial region of sensitivity of the potentials 
\cite{brandan,tenreiro,pak3,roubos},{\it i.e.} the region where the potentials are best determined by the elastic 
scattering data. In order to find the radius of sensitivity or the {\it crossing radius}, we adopted the procedure 
described in Ref.\cite{woj}. At each energy, we changed the diffuseness parameter {\it a} in small steps around the
best fit value and adjusted the strength and radius values of the potential to fit the angular distribution data. 
The process resulted into a family of 'good' potentials having $\chi^2$ = ${\chi^2}_{min}$ + 1.
Almost equivalent reproduction of the angular distribution and the crossing of the potentials at a certain radius value, 
the {\it crossing radius}, ensure that the crossing point gives the value of the sensitive radius. Same procedure
was adopted to obtain both the real and the imaginary sensitive radial region. 
Both the real and imaginary sensitive radii were found to be energy dependent.
The crossing radius moved to lower radius value as the incident energy was
increased. The average value of the crossing radius was estimated to be
9.8 $\pm$ 0.4 fm. The strengths of the potential components were determined at
this radius to obtain the energy dependent behaviour of the potentials. In
Fig.2, the values at the chosen radius of the potentials with Woods Saxon form are shown with
solid circles. The error bar with each point includes the distribution in the
values of the different 'good' potentials at the chosen radius as well as the errors due to
the uncertainty associated with the sensitive radius value itself.

\subsection{OM analysis of $^6$Li+$^{64,58}$Ni data with folded potential}

Following the analysis of the $^6$Li+$^{64}$Ni data with the phenomenological Woods Saxon potential, 
a further analysis of the new data of $^6$Li+$^{64}$Ni and the existing data of $^6$Li+$^{58}$Ni 
\cite{pfeiffer} were performed with folded M3Y potential. Alongwith the real folded potential, two 
different imagianry model potentials were used. In the first one (OMP1) the model potential had the 
form

\begin{equation}
U_{OM}(R) = \lambda_rV_f(R)+i\lambda_iV_f(R).
\end{equation}

where $V_f$ is the folded M3Y potential with unit normalization. $\lambda_r$ and $\lambda_i$ are the renormalization 
factors for real and imaginary components. With geometries fixed, search was performed on these two strength parameters 
simultaneously to fit the angular distributions. 

The second model potential (OMP2) was of {\it hybrid} nature with the double-folded real potential 
and the phenomenological 
imaginary potential 

\begin{equation}
U_{OM}(R) = \lambda_rV_f(R)+iW_v(R;W_o,r_w,a_w)
\end{equation}

where W$_v$ is the imaginary potential of volume Woods-Saxon type. The OMP2 with fixed geometry real part, therefore, has 
four parameters, {\it viz.} $\lambda_r$, W$_o$, r$_w$, a$_w$ to vary to fit the experimental angular distributions.  

The double-folded real potentials were generated using the density dependent M3Y-Reid nucleon-nucleon interaction 
with zero-range exchange term \cite{khoa}. The density dependence was included following the DDM3Y convention \cite{kobos} 
as

\begin{equation}
F(\rho)= C[ 1 + \alpha exp(-\beta\rho) ]
\end{equation}
 
with the parameters C, $\alpha$ and $\beta$ having the values 0.2845, 3.6391 and 2.9605 $fm^3$ respectively. A linear 
energy dependent part g(E)=(1-0.002E) had been considered \cite{khoa} to account for explicit energy dependence of the 
effective interaction. The mass densities of $^{58}$Ni and $^{64}$Ni were obtained from Ref.\cite{ripl2}. A parametric 
form for $^6$Li charge density \cite{satch1} was unfolded for finite proton distribution to obtain the point proton 
distribution. The neutron distribution was assumed to have the same radial shape for N=Z $^6$Li nucleus. 

With the model potential OMP1, the two free parameters
$\lambda_r$ and $\lambda_i$ were varied simultaneously to optimize the fits to the data at all the energies.
The best fit values of  $\lambda_r$ and $\lambda_i$ are given in Table 2 and the corresponding fits are 
shown by dotted lines in Figs.1 and 3 for $^6$Li+$^{64}$Ni and $^6$Li+$^{58}$Ni respectively.  The same condition, as described in the case of phenomenological
potential, was used to determine the error limits in the values of the individual renormalization factors.
To obtain the best fit parameters of the OMP2 potentials for both the isotopes, analysis was started with the
highest energy data.  At the highest energy, we performed an initial search over all the
four parameters simultaneously. Subsequently, the imaginary radius parameter obtained from the initial search was 
kept fixed. The best fit, determined by $\chi^2$ minimization, was found by searching over the real renormalization 
factor and the imaginary strength while gridding over the imaginary diffuseness a$_w$. Same search procedure was adopted 
for all the incident energies. The radius parameter was held fixed throughout assuming that the change in the value of
r$_w$ due to change in incident energy is not so significant. The best fit parameters along with the $\chi^2$/N ( N
denotes the number of data points) values and the reaction cross sections $\sigma_{reac.}$ have been given in Table 2.
The errors associated with the strength parameters and the reaction cross sections were estimated in the same way as
described earlier. The best fit predictions with OMP2 for $^6$Li+$^{64}$Ni are shown in Fig.1 by dashed-dotted line and 
for $^6$Li+$^{58}$Ni are shown by solid lines in Fig.3. 

In order to find the radius of sensitivity or the {\it crossing radius} with potential OMP2,  
we searched over $\lambda_r$ and $W_0$ at different diffuseness parameter $a_w$, obtained 
by changing its value in small steps while keeping the radius fixed. The $\chi^2$ minimization process yielded different 
sets of parameter values for OMP2 producing almost equivalent fits to the data. The $\chi^2_{min}$ value for each set of 
parameters lies within the range of ${\chi^2_{min}\pm\chi^2_{min}}/N$ corresponding to the set of best fit parameters. 
In Fig.4 the crossing 
points at 14 MeV laboratory energy for $^6$Li+$^{58}$Ni and $^6$Li+$^{64}$Ni systems have been compared. It is to be noted 
that with OMP2 we have obtained the sensitive radius of the imaginary potential only. The observed crossing points are 
quite sharp for $^6$Li+$^{64}$Ni but for $^6$Li+$^{58}$Ni the crossing points are not so unambiguos at certain energies. 
The radii of sensitivity for imaginary potentials found from the analyses of 
the data vary with the incident energy. The value increases by $\sim$25\% at the lowest energy studied. At the higher 
incident energies the values of the crossing radii are close to the strong absorption radii, R$_{sa}$, of the systems. 

Considering the energy variation of the crossing radius and the uncertainty associated with it, 
we have used in the dispersion relation the integral quantities G$_x$(E) defined as  

\begin{equation}
G_x(E) = \frac{4\pi}{A_PA_T} \int{x(R,E)G(R)R^2dR}
\end{equation}

where G(R) is the Gaussian weighting function centered on some average value of the crossing radii and {\it x} denoting
either the real potential V(R,E) or the imaginary potential W(R,E) for the respective integral quantity. It was 
demonstrated that the dispersion relation can be extended to the integral quantities with Gaussian weighting function 
\cite {mahaux,mahaux1}, such that

\begin{equation}
G_V(E) = G_{V_o}(E)+\Delta G_V(E)
\end{equation}
where

\begin{equation}
\Delta G_V(E) = \frac{P}{\pi}\int{\frac{G_W(E')}{E'-E} dE'}
\end{equation}

The Gaussian weighting function has the usual form of 

\begin{equation}
G(R) = \frac{1}{\sqrt{2\pi}\sigma_G} exp{ \large[-\frac{(R-R_G)^2}{2\sigma_G^2} ]}
\end{equation}

The choices of radius parameter $R_G$ and the width parameter $\sigma_G$ of the Gaussian weighting function
are very crucial for systems involving light, loosely bound projectiles. Although the regions of sensitivity 
are narrow at each energy, the crossing radii vary with energy. To identify the suitable values of R$_G$
and $\sigma_G$, we estimated the uncertainties associated with the integral quantities with different combinations of
(R$_G$,$\sigma_G$). It can easily be understood that the range of uncertainties associated with G$_V$(E)
and G$_W$(E) at a particular energy changes significantly depending upon the proximity of the chosen R$_G$ to
crossing radius at that energy. So the final
choice of R$_G$ and $\sigma_G$ is a compromise relative to the uncertainties associated with G$_V$(E) and G$_W$(E) at
each energy. The chosen value of R$_G$ is 9.8fm for both $^{64}$Ni and $^{58}$Ni targets.
The associated $\sigma_G$ was chosen to be 0.6fm. We would like to emphasize that the choice of R$_G$ does not really 
change the overall trend of the energy variations of G$_V$(E) and G$_W$(E). The quantities G$_V$(E) and G$_W$(E)
as functions of incident energy for $^{64}$Ni and $^{58}$Ni are shown in Figs.5 and 6. The error bars shown in the figures
predominantly arise out of the variations in the values of the {\it good potentials} at R$_G$ within the width defined 
by $\sigma_G$.  
In Fig.7, we have shown the best fit normalizations $\lambda_r$ and $\lambda_i$, obtained with model 
potential OMP1, as a function of energy for $^6$Li+$^{58,64}$Ni. Errors shown are estimated individually for each 
parameter using the condition of Eq.4. The plotting of these normalization factors directly is justified considering 
the fact that dispersion relation involving these factors will not depend on the choice of the radial point for the 
evaluation of the energy integration.

\section{Results and Discussion}

The phenomenological analysis of $^6$Li+$^{64}$Ni data with Woods Saxon potential having variable geometry with 
decreasing energy yielded an energy dependence of the surface strengths of the real and imaginary components shown
in Fig.2. The variations resulted from the model potentials OMP1, with fixed geometry real and imaginary components, 
and OMP2, with fixed geometry real and variable geometry imaginary components, are also shown in Fig.2. Good matching
in the surface strengths of the imaginary potential at all the energies have been obtained for the three different
model potentials used in the anlysis. The magnitude of the real strength, however, shows some kind of model dependence.
Despite the difference in the magnitudes, the three model potentials exhibit similar trend in the energy variation
of the real strength. The observed variation indicates that the strength of the imaginary 
potential increases initially as the $^6$Li projectile energy approaches the barrier for the system.   
With further decrease of incident energy the strength of the imaginary potential starts to decrease. 
On the otherhand, the real strength shows an overall increase with decreasing energy although there is a tendency
to decline around 18 MeV energy. This feature is more prominent in the case of phenomenological potential with
variable geometry.

The observed variations of G$_W$(E) in Figs.5 and 6 also indicate the same behaviour for the strengths of the imaginary 
and the real potentials for both the Ni isotopes as the $^6$Li projectile energy approaches the barrier for the systems.   
In order to find the connectivity between the energy dependence of the real and imaginary components through the 
dispersion relation, we fitted the observed variation of G$_W$(E) with the following functional form

\begin{eqnarray}
G_W(E)      = & 0 &~~~  ,~ E \leq E_1 \nonumber \\
            = & C_1 (E-E_1) &~~~ ,~E_1 \leq E \leq E_2 \nonumber \\
            = & C_1 (E_2-E_1)+C_2(E-E_2) &~~~ ,~E_2 \leq E \leq E_3 \nonumber \\
            = & C_3 &~~~ ,~E \geq E_3
\end{eqnarray}

With this three linear segments form of energy dependence for G$_W$(E), the dynamical contribution to the real part, 
{\it i.e.} the real polarization component is given by

\begin{eqnarray}
\Delta G_V(E) & = & \frac{1}{\pi} [(G_W(E_1)-G_W(E_2))(\epsilon_1ln|{\epsilon_1}|-\epsilon_2ln|\epsilon_2|) \nonumber \\
 &  &+(G_W(E_2)-G_W(E_3))(\epsilon'_2 ln|{\epsilon'_2}| - \epsilon'_3 ln|{\epsilon'_3}|)]
\end{eqnarray}

where  $\epsilon_i$=$(E-E_i)/(E_2-E_1)$ and ${\epsilon'}_i$=${(E-E_i)/(E_3-E_2)}$. The energy $E_1$ is the threshold energy
where the absoprtion goes to zero. It has been determined assuming that the function ($\sqrt (E \sigma_{reac.})$) behaves 
as a linear function of energy E in the sub-barrier region \cite{stelson}. $E_2$ is the energy at which the increasing and 
the decreasing linear segments intersect while $E_3$ is the energy beyond which the variation of $G_W$(E) or rather the 
imaginary potential W(E) is assumed to be independent of energy. The values of the coefficients $C_i$ and energies $E_i$ 
for both the systems are given in Table 2. The solid curves in the plots of $G_V$(E) vs. E in Figs.5 \& 6 indicate the 
dispersion relation prediction following Eq.10. The general agreement with the observed trend in $G_V$(E) is quite good 
considering the simple form employed to describe $G_W$(E).

As it appears from Figs.5 and 6, the imaginary potential, for both $^6$Li+$^{58}$Ni and $^6$Li+$^{64}$Ni, shows a 
distinct increasing trend as the energy approaches the barrier but with further lowering of energy below the barrier 
it starts to decrease. The interplay of the dispersive contributions of these two regions in W(E) constitute the 
effective energy dependence of the real potential. The resultant energy dependence of the real polarization 
potential exhibits an overall increasing trend. This suggests to the presence of {\it thershold anomaly} (TA) for
$^6$Li scattering from medium mass Ni isotopes. However, a careful inspection also reveals that as the bombarding 
energy decreases and approaches the barrier, the strength of the real effective potential
increases slowly at the beginning and then shows a weak declining trend around 1.2 times the barrier before sharply 
rising again. This decreasing tendency of the real potential is associated with the increasing part of the imaginary
potential. Similar behaviour for the real and imaginary components of the effective potential has also been observed 
for $^6$Li+$^{208}$Pb \cite{hussein}, $^6$Li+$^{59}$Co \cite{beck2, souza} and $^6$Li+$^{27}$Al \cite{ferna,figu} 
systems and has been termd as {\it breakup threshold anomaly} or BTA. For $^6$Li+$^{208}$Pb \cite{hussein}, the 
strength of the real effective potential continues to decrease even below the barrier indicating the presence of 
reaction channels, like break up, producing strong repulsive real polarization even at sub barrier energies. However, 
our investigation of $^6$Li scattering from medium mass Ni-isotopes suggests that the observed threshold anomaly has 
the usual form albeit modified by coupling to breakup (or channels producing repulsive real polarization). A recent 
two-body continuum discretized coupled channel (CDCC) calculation \cite{beck2} for $^6$Li on $^{59}$Co has shown that 
the real potential does exhibit an overall increasing trend while the imaginary potential falls of after a short 
increase near the barrier. It is, therefore, extremely important to investigate the target mass dependence of 
{\it threshold anomaly} for projectiles with low breakup threshold, as has been pointed out by Beck, {\it et al.} 
\cite{beck2}, in order to understand the evolution of breakup coupling with increasing or decreasing mass at near barrier
energies. 

In Fig.7a) \& b) we have compared the observed energy dependence of the renormalization parameters $\lambda_r$ and 
$\lambda_i$ obtained by fitting the elastic angular distributions of $^6$Li+$^{58,64}$Ni with potential OMP1. It is 
to be noted that the variation of the renormalization parameters is independent of the choice of interaction distance.  
For both the systems the variations in $\lambda_r$ and $\lambda_i$ with incident energy are very similar to the 
variations observed in the integral quatities $G_V$(E) and $G_W$(E). The observed energy dependence of $\lambda_r$ 
and $\lambda_i$ for $^6$Li+$^{58}$Ni and $^6$Li+$^{64}$Ni in Fig.7 does not indicate of any dramatic isotopic variation. 
However, careful observation will show that the rise in the imaginary strength for the $^6$Li+$^{64}$Ni system starts 
earlier than the $^6$Li+$^{58}$Ni system. This is also obvious from the values of E$_2$ parameter in Table 2. The 
the fall off in the imaginary strength around the barrier is slower in the case of lighter Ni isotope but the large error
bars in the lowest energies do not allow the observation to be conclusive. Exclusive coincident measurement of fragments 
produced in break up or transfer followed by break up reactions at these bombarding energies can throw some light on the 
possible difference in the observed behaviour. \\

\section{Summary}

In this work the elastic scattering angular distributions measured in the bombarding energy range 13 MeV 
$\leq E_{lab} \leq$ 26 MeV around the Coulomb barrier for the system $^6$Li+$^{64}$Ni have been presented.
The measured elastic scattering data for $^6$Li+$^{64}$Ni has been investigated with three different model 
potentials - a phenomenological Woods Saxon potential with variable geometry as energy changes, a fixed 
geometry folded DDM3Y potential with complex renormalization factor and a {\it hybrid} potential with fixed geometry
real folded potential and variable geometry phenomenological imaginary potential. The existing data for the 
$^6$Li+$^{58}$Ni system have, subsequently, been analyzed using the folded DDM3Y potential with complex normalization
and also the hybrid potential model. The fitting procedure yielded the energy dependence of the real and imaginary
components - the data required to demonstrate the existence or non-existence of the threshold anomaly for 
$^6$Li+$^{58}$Ni and $^6$Li+$^{64}$Ni systems.

The near threshold behaviour of all the model interaction potentials are similar in nature for both the systems. 
The imaginary potential falls of after a short increasing region close to the barrier. On the otherhand, the real
potential shows an overall increasing trend with decreasing energy. But there is definite tendency to decline in
the energy region where the imaginary potential shows an increasing nature.  
The observed energy dependences of the real and imaginary potentials corroborate with the observations from the study
of $^6$Li+$^{59}$Co and $^6$Li+$^{27}$Al systems. This particular nature of the interaction potential close to the 
barrier has been argued to be due to the coupling to the continuum. However, unlike the $^6$Li+$^{208}$Pb system, 
the real potential instead of decreasing further in the sub-barrier energy region shows a sharp rise. A dispersion
relation prediction for the real potential behaviour assuming a simple 3-linear segment description of the energy
dependence of the imaginary potential describes nicely the observed energy variation of the real component.  
In order to pinpiont the threshold behaviour of the interaction potential with loosely bound projectiles, further
experimental investigations with more precision and in smaller energy steps around the barrier is necessary. Also
a complete CRC calculation including the coupling to the inelastic and other 
direct reaction channels in the presence of coupling to the continuum is necessary to understand the observed energy
variations of the real and imaginary potentials for $^6$Li+$^{58}$Ni and $^6$Li+$^{64}$Ni systems.

\begin{table}
\setlength{\tabcolsep}{0.06in} 
\caption{\label{tab:Table1}Best fit parameters with phenomenological potential (PH) for $^6$Li+$^{64}$Ni}
\begin{tabular}{|c|c|c|c|c|c|c|c|c|}\hline
$E_{lab}$(MeV)& $V_0$ & $R_0$ & $a_0$ & $W_s$ & $R_s$ & $a_s$ & $\chi^2$/N & $\sigma_R$(mb)\\
\hline
13.0 &34.1 &6.736 &0.737 & 5.107 & 7.196 & 0.703 & 0.13 & 196.5\\
14.0 &24.4 &6.457 &0.760 & 4.081 & 7.196 & 0.812 & 0.54 & 388.4\\
17.0 &38.1 &5.978 &0.743 & 3.18 & 7.218 & 0.838 & 0.40 & 813.1\\
19.0 &36.8 &5.936 &0.755 & 3.743 & 6.944 & 0.844 & 0.45 & 983.2\\
26.0 &46.0 &5.860 &0.757 & 7.73 & 6.000 & 0.944 & 0.45 & 1538.9\\
\hline
\end{tabular}
\end{table}

\begin{table}
\setlength{\tabcolsep}{0.03in} 
\caption{\label{tab:Table2}Best fit potential parameters with folded potential }
\begin{tabular}{|c|c|c|c|c|c|c|c|c|}\hline
$E_{lab}$(MeV)&Model& $\lambda_r$ & $\lambda_i$ & $W_0$ & $R_w$ & $a_w$ & $\chi^2$/N & $\sigma_R$(mb)\\
\hline
\multicolumn{9}{|c|}{For $^6$Li+$^{64}$Ni}\\
\hline
13.0 &OMP1& 0.88$\pm$0.21&0.91$\pm$0.3&&&&0.13&196.0$\pm$29.0 \\
&OMP2& 0.93$\pm$0.11&& 52.01$\pm$18.00 & 6.753 & 0.665 & 0.13 & 197.8$\pm$10.8\\
14.0 &OMP1&0.75$\pm$0.05&1.240$\pm$0.11&&&&0.52&372.25$\pm$11.6 \\
&OMP2& 0.75$\pm$0.04&& 41.13$\pm$3.67 & 6.753 & 0.724 & 0.58 & 369.9$\pm$11.7\\
17.0 &OMP1& 0.69$\pm$0.02&0.88$\pm$0.02&&&&0.43&765.8$\pm$6.0 \\
&OMP2& 0.60$\pm$0.02&& 31.92$\pm$1.60 & 6.753 & 0.744 & 0.40 & 778.3$\pm$9.1\\
19.0 &OMP1& 0.60$\pm$0.02&0.72$\pm$0.05&&&&0.54&945.7$\pm$9.8 \\
&OMP2& 0.51$\pm$0.02&& 23.56$\pm$3.70 & 6.753 & 0.764 & 0.53 & 953.4$\pm$15.2\\
26.0 &OMP1& 0.63$\pm$0.04&0.61$\pm$0.04&&&&0.47&1407.8$\pm$20.57 \\
&OMP2& 0.60$\pm$0.06&& 23.87$\pm$2.93 & 6.753 & 0.766 & 0.53 & 1460.3$\pm$38.8\\
\hline
\multicolumn{9}{|c|}{For $^6$Li+$^{58}$Ni}\\ 
\hline
12.0 &OMP1&0.99$\pm$0.09&0.90$\pm$0.13&&&&0.11&52.5$\pm$5.1\\
&OMP2&1.15$\pm$0.21&& 97.84$\pm$41.00 & 6.598 & 0.585 & 0.11 & 51.9$\pm$12.0\\
14.0 &OMP1&0.68$\pm$0.02&1.09$\pm$0.06&&&&0.18&269.5$\pm$6.0 \\
&OMP2&0.81$\pm$0.03&& 70.85$\pm$11.01 & 6.598 & 0.633 & 0.18 & 266.3$\pm$14.4\\
16.0 &OMP1&0.64$\pm$0.02&0.90$\pm$0.02&&&&0.51&528.6$\pm$1.5 \\
&OMP2&0.66$\pm$0.02&& 39.53$\pm$4.48 & 6.598 & 0.686 & 0.52 & 526.7$\pm$15.3\\
18.0 &OMP1&0.55$\pm$0.004&0.63$\pm$0.01&&&&0.25&709.7$\pm$2.2 \\
&OMP2&0.42$\pm$0.014&& 18.77$\pm$2370 & 6.598 & 0.795 & 0.23 & 746.4$\pm$24.3\\
20.0 &OMP2&0.57$\pm$0.01&0.69$\pm$0.02&&&&1.5&922.5$\pm$0.2 \\
&OMP2&0.45$\pm$0.02&& 22.12$\pm$1.70 & 6.598 & 0.797 & 0.94 & 980.6$\pm$17.6 \\ \hline
\end{tabular}
\end{table}

\begin{table}
\setlength{\tabcolsep}{0.10in} 
\caption{\label{tab:Table3}Parameters of 3-linear segment fit of $G(E)_W$ }
\begin{tabular}{|c|c|c|c|c|c|} \hline
$E_1$(MeV) & $E_2$(MeV) & $E_3$(MeV) & $C_1$ & $C_2$ & $C_3$ \\
\hline
\multicolumn{6}{|c|}{For $^6$Li+$^{64}$Ni} \\
\hline
9.56&13.49&19.0&0.639&-0.156&1.62 \\
\hline
\multicolumn{6}{|c|}{For $^6$Li+$^{58}$Ni} \\
\hline
10.03&12.25&17.78&1.09&-0.192&1.36 \\ \hline
\end{tabular}
\end{table}

\vspace{+5.5cm}

\section{Acknowledgments}
                                                                                
The authors thank the BARC/TIFR Pelletron staff for delivering the $^6$Li beam. Thanks are also due 
to Mr. Pradipta Das and Mr. B.P. Das for their help in preparing the targets. We also thank 
Mr. Ajay K. Mitra for his extensive support during the experiment.

\newpage
\begin{center}
{\bf 	FIGURE CAPTIONS}
\end{center}

\begin{itemize}
\item{Fig. 1. }{Elastic angular distributions of $^6$Li+$^{64}$Ni.
The solid line represents prediction with the phenomenological potential. The dashed-dotted (dotted) curves
are the predictions using the model potential OMP2 (OMP1).}

\item{Fig. 2. }{The energy dependence of the real and imaginary surface strengths at an average radius 
of 9.8 fm obtained with the phenomenological (solid circle), OMP1 (open circle) and OMP2 (solid triangle) 
model potentials for $^6$Li+$^{64}$Ni system.}

\item{Fig. 3.} {Elastic angular distributions of $^6$Li+$^{58}$Ni.
The solid (dotted) line represents prediction of OMP2 (OMP1).}

\item{Fig. 4.} { Crossing points of 'good' imaginary potentials (OMP2) at 14 MeV for
$^6$Li+$^{58}$Ni and $^6$Li+$^{64}$Ni systems. }

\item{Fig. 5.} { Energy  variations of Gauss weighted integral quantities of real and 
imaginary components of OMP2 for $^6$Li+$^{64}$Ni. The solid line in the upper panel is
the dispersion relation prediction with 3-linear segment fit of the imaginary component.
The dotted line in the upper panel depicts the intrinsic energy dependence of the Gauss 
weighted integral of unrenormalized real part of OMP2. $E_{C.b.}$=13.8 MeV is the value of the 
Coulomb barrier for $^6$Li+$^{64}$Ni system in the laboratory \cite{browin}}

\item{Fig. 6.} { Energy  variations of Gauss weighted integral quantities of real and 
imaginary components of OMP2 for $^6$Li+$^{58}$Ni. The solid line in the upper panel is
the dispersion relation prediction with 3-linear segment fit of the imaginary component. 
The dotted line in the upper panel depicts the intrinsic energy dependence of the Gauss 
weighted integral of unrenormalized real part of OMP2. $E_{C.b.}$=14.1 MeV is the value of the 
Coulomb barrier for $^6$Li+$^{58}$Ni system in the laboratory \cite{browin}}

\item{Fig. 7.} { Energy dependence of the effective real and imaginary normalization factors
with model potential OMP1. The solid circle corresponds to $^6$Li+$^{64}$Ni and the hollow
circle to $^6$Li+$^{58}$Ni. $E_{C.b.}$ for the two systems are specified in the captions of 
Figs. 5 and 6. }   
\end{itemize}

\newpage
\begin{figure}
\vspace{16.5cm}
\includegraphics{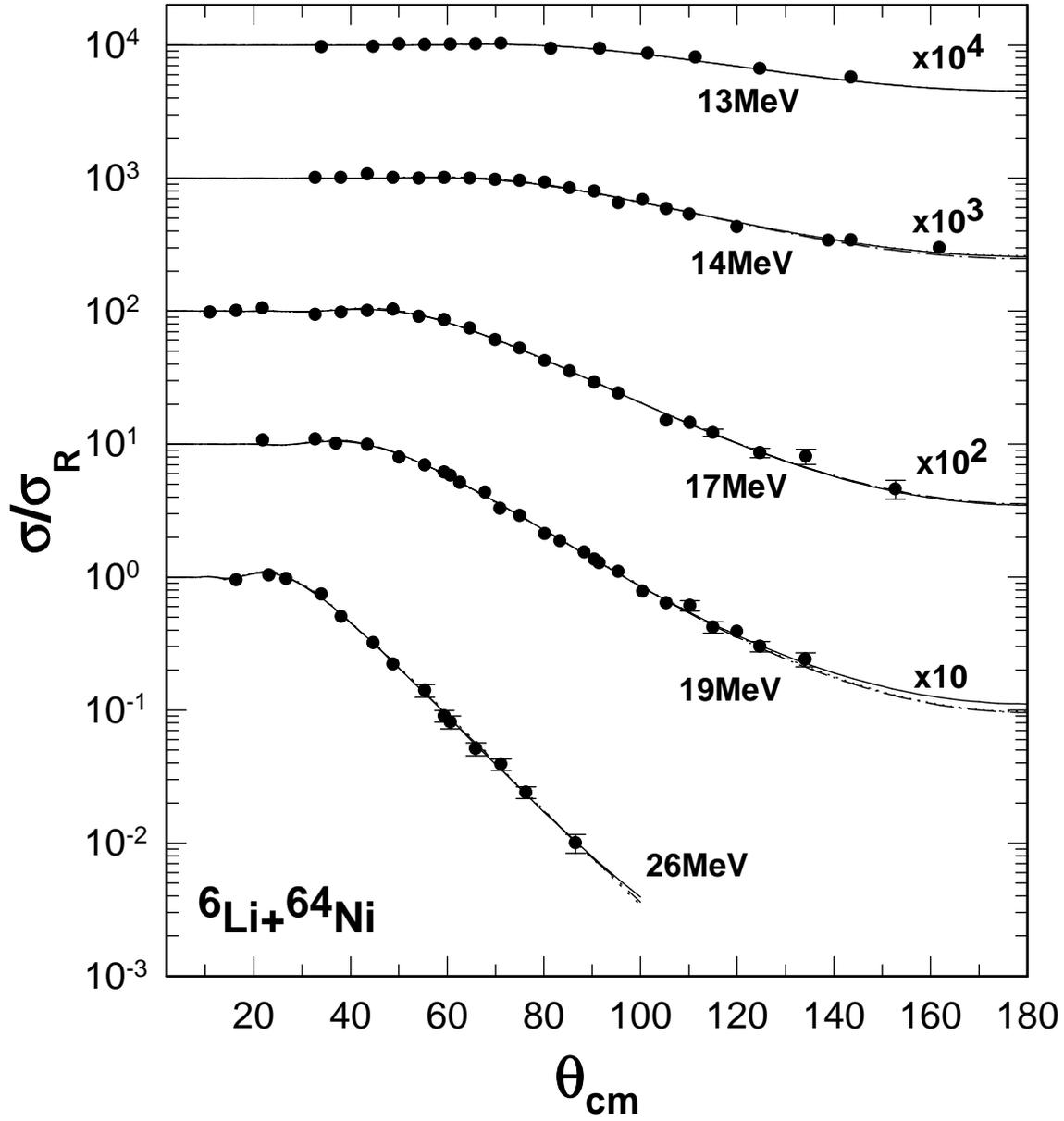}
\caption {\label{fig:fig1}Elastic angular distributions of $^6$Li+$^{64}$Ni.
The solid line represents prediction with the phenomenological potential. The dashed-dotted (dotted) curves
are the predictions using the model potential OMP2 (OMP1).}
\end{figure}

\begin{figure}
\vspace{16.5cm}
\includegraphics{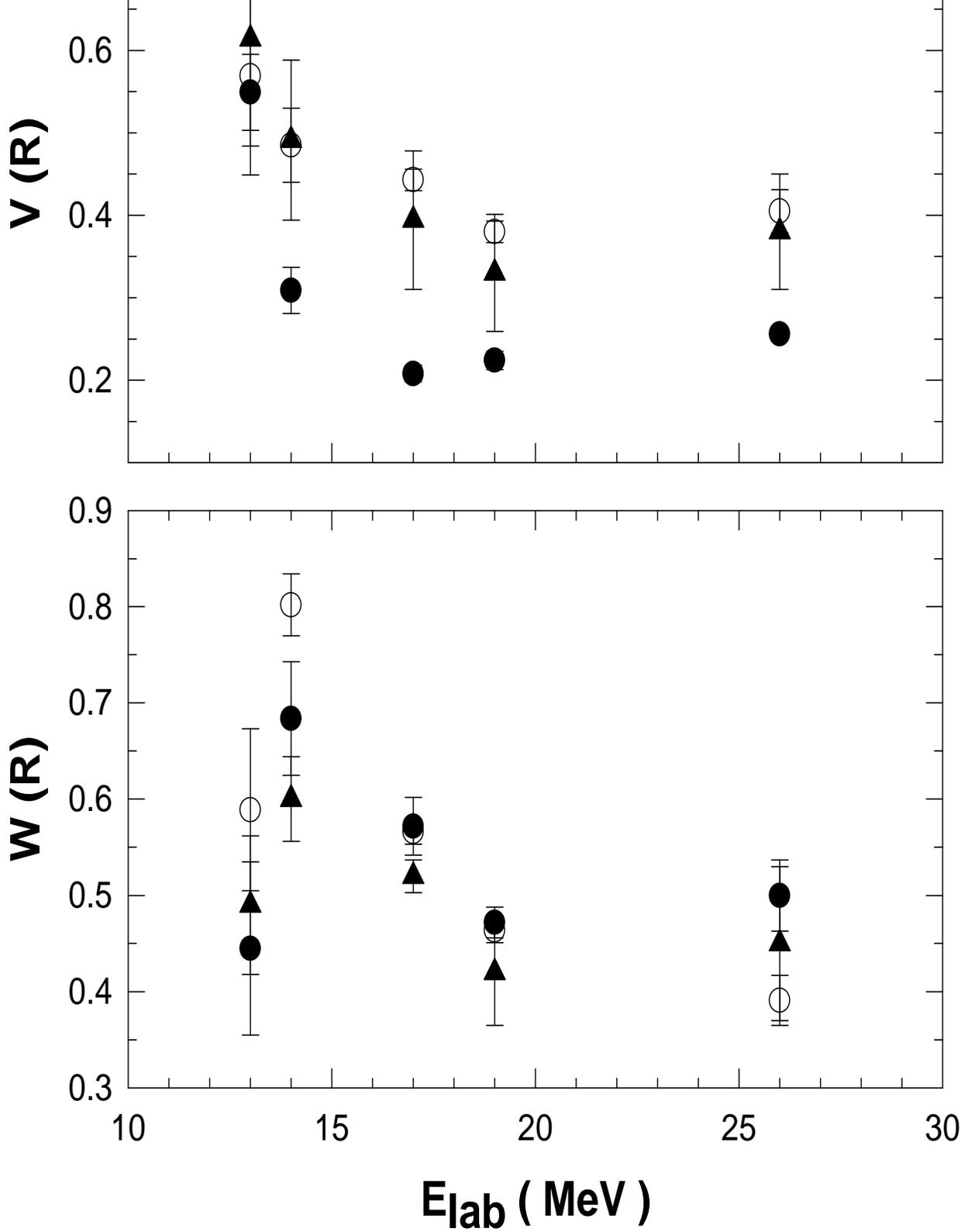}
\caption {\label{fig:fig2} The energy dependence of the real and imaginary surface strengths at an average radius 
of 9.8 fm obtained with the phenomenological (solid circle), OMP1 (open circle) and OMP2 (solid triangle) model 
potentials for $^6$Li+$^{64}$Ni system.}
\end{figure}

\begin{figure}
\vspace{16.5cm}
\includegraphics{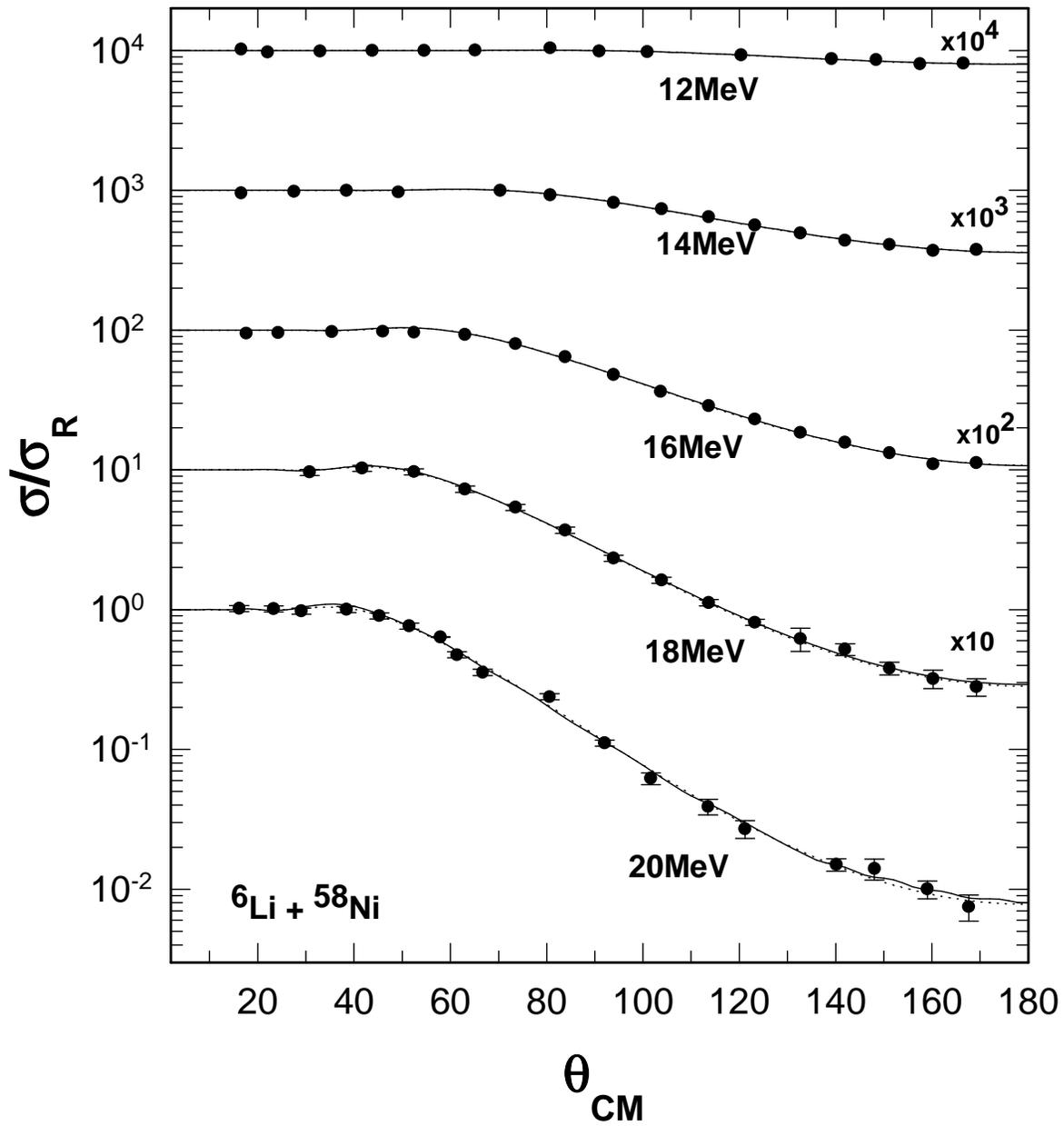}
\caption {\label{fig:fig3}Elastic angular distributions of $^6$Li+$^{58}$Ni.
the solid (dotted) line represents prediction of OMP2 (OMP1).}
\end{figure}

\begin{figure}
\vspace{16.5cm}
\includegraphics{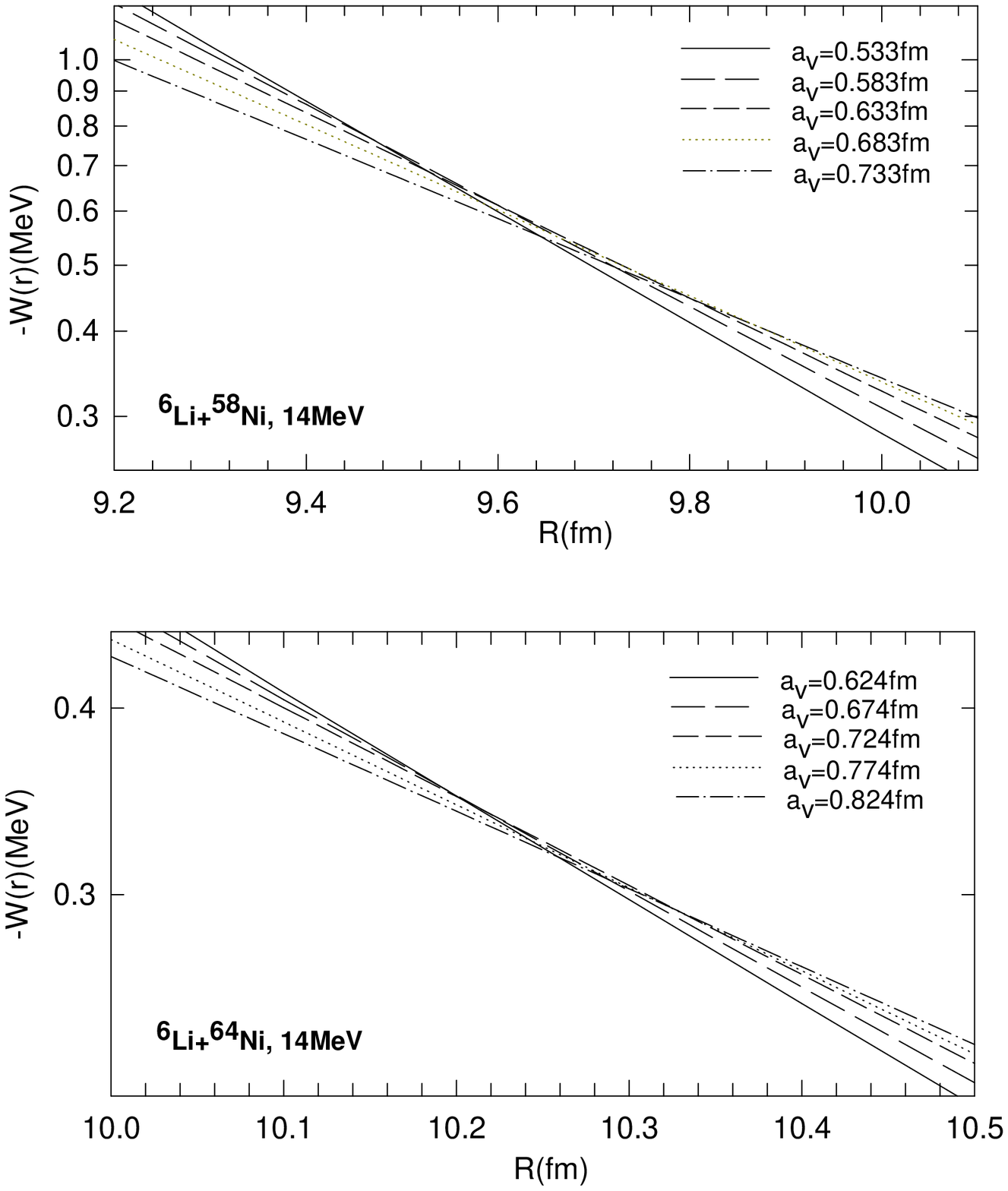}
\caption {\label{fig:fig4}Crossing points of 'good' imaginary potentials (OMP2) at 14 MeV for
$^6$Li+$^{58}$Ni and $^6$Li+$^{64}$Ni systems.}
\end{figure}

\begin{figure}  
\vspace{16.5cm}
\includegraphics{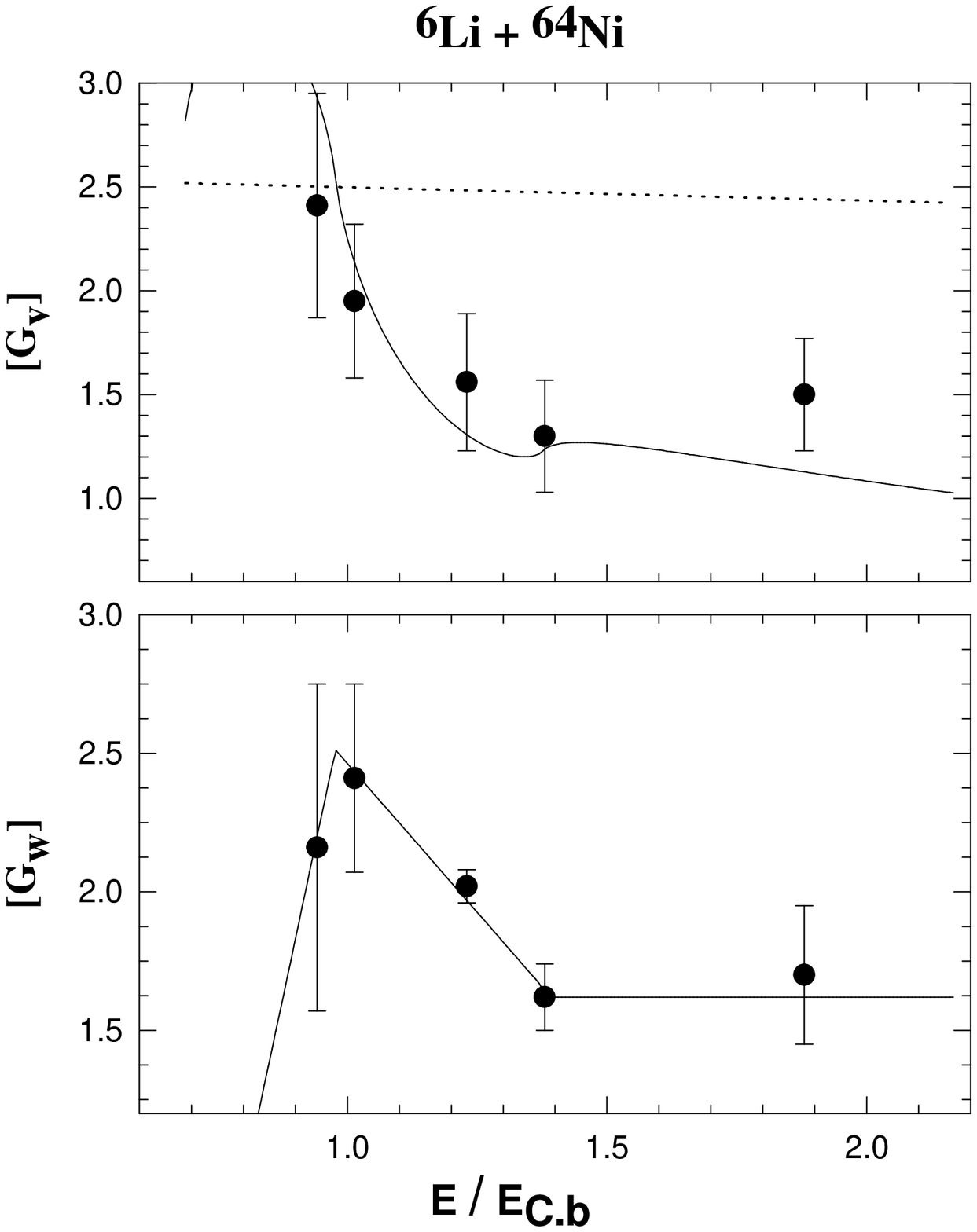}
\caption {\label{fig:fig5}Energy  variations of Gauss weighted integral quantities of real and 
imaginary components of OMP2 for $^6$Li+$^{64}$Ni. The solid line in the upper panel is
the dispersion relation prediction with 3-linear segment fit of the imaginary component.
The dotted line in the upper panel depicts the intrinsic energy dependence of the Gauss 
weighted integral of unrenormalized real part of OMP2. $E_{C.b}$=13.8 MeV is the value of the 
Coulomb barrier for $^6$Li+$^{64}$Ni system in the laboratory \cite{browin} }
\end{figure}

\begin{figure}[h]
\vspace{16.5cm}
\includegraphics{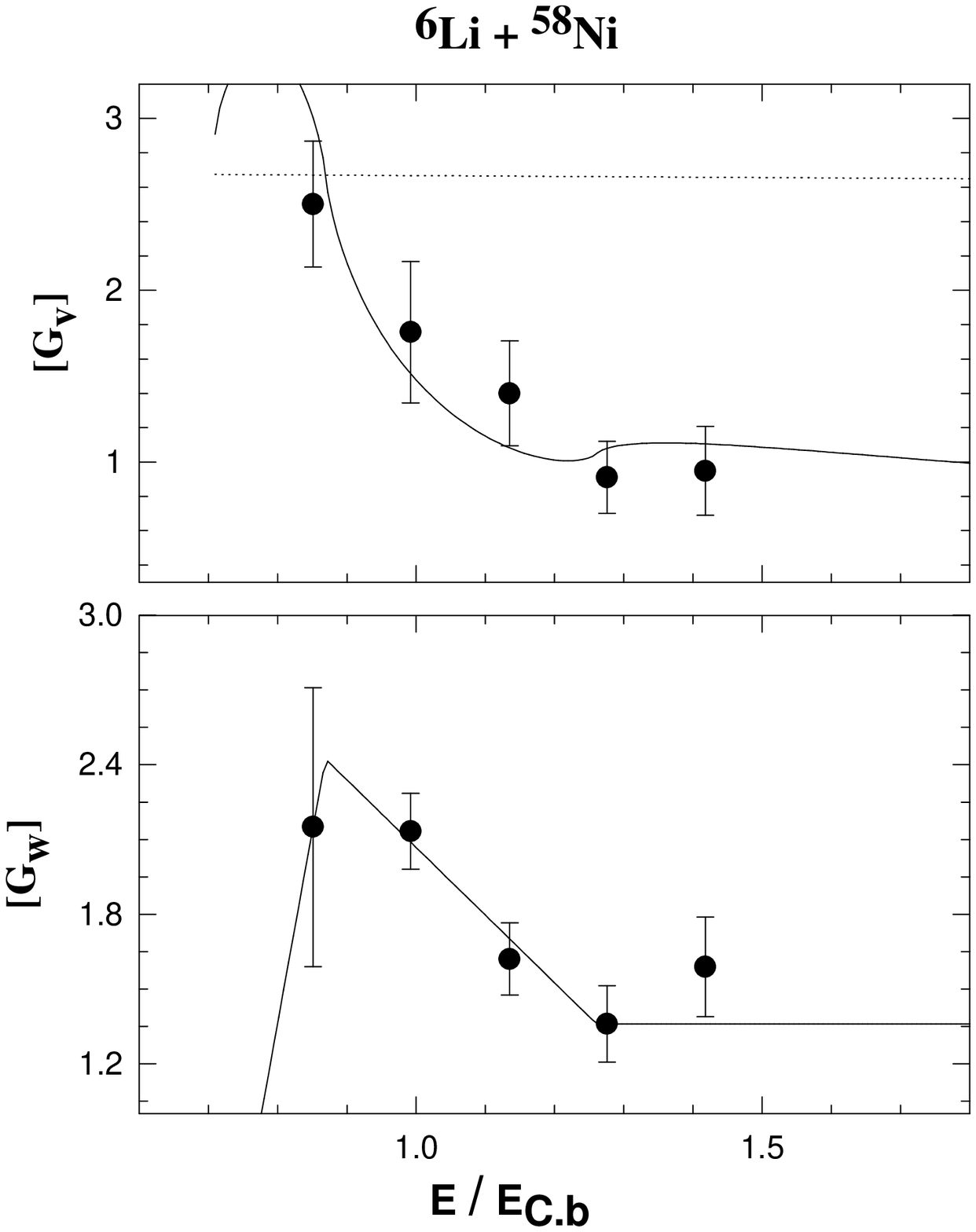}
\caption{\label{fig:fig6}Energy  variations of Gauss weighted integral quantities of real and 
imaginary components of OMP2 for $^6$Li+$^{58}$Ni. The solid line in the upper panel is
the dispersion relation prediction with 3-linear segment fit of the imaginary component. 
The dotted line in the upper panel depicts the intrinsic energy dependence of the Gauss 
weighted integral of unrenormalized real part of OMP2. $E_{C.b}$=14.1 MeV is the value of the 
Coulomb barrier for $^6$Li+$^{58}$Ni system in the laboratory \cite{browin}}
\end{figure}

\begin{figure}[h]
\vspace{16.5cm}
\includegraphics{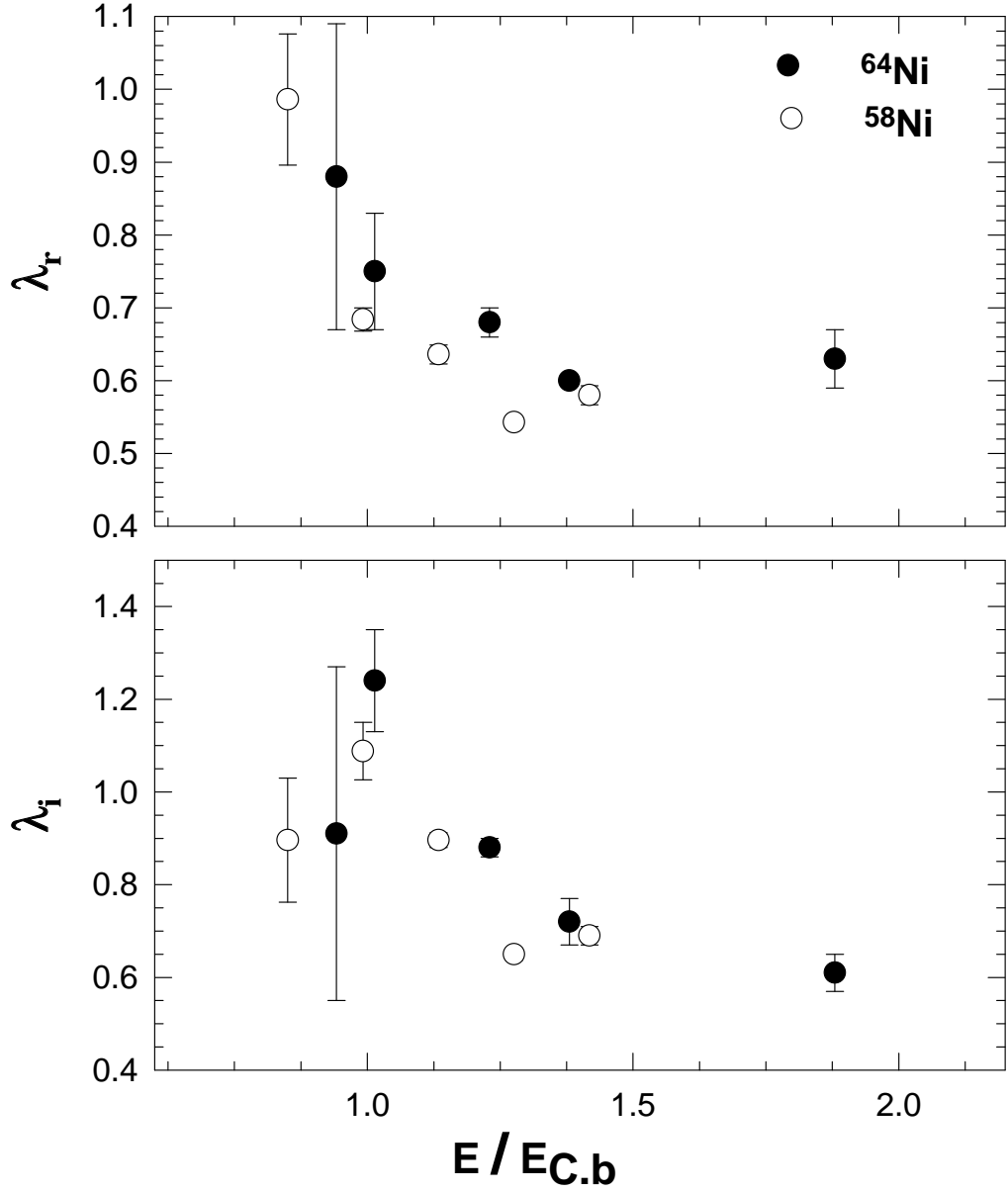}
\caption{\label{fig:fig7}Energy dependence of the effective real and imaginary normalization factors
model potential OMP1. The solid circle corresponds to $^6$Li+$^{64}$Ni and the hollow
circle to $^6$Li+$^{58}$Ni. $E_{C.b}$ for the two systems are specified in the captions of 
Figs. 5 and 6. } 
\end{figure}

\end{document}